\documentstyle[12pt,a4wide]{article}

\newcommand{\ssf}{\sans}
\newcommand{\news}{\setcounter{equation}{0}}
\newcommand{\be}{\begin{equation}}
\newcommand{\ee}{\end{equation}}
\newcommand{\R}{\hbox{\upright\rlap{I}\kern 1.7pt R}}
\newcommand{\Z}{\hbox{\upright\rlap{\ssf Z}\kern 2.7pt {\ssf Z}}}
\newcommand{\bea}{\begin{eqnarray}}
\newcommand{\eea}{\end{eqnarray}}
\newcommand{\alphabf}{\mbox{\boldmath$\alpha$}}
\newcommand{\betabf}{\mbox{\boldmath$\beta$}}
\newcommand{\gammabf}{\mbox{\boldmath$\gamma$}}

\font\upright=cmu10 scaled\magstep1
\font\sans=cmss12

\begin{document}
\pagestyle{plain}
\title{\vskip -70pt
  \begin{flushright}
    {\normalsize{
    UDEM-GPP-TH 00-67.\\ hep-th/0004054.}} 
  \end{flushright}
  \vskip 15pt
  {\bf \Large \bf Radiation from SU(3) monopole scattering.} 
  \vskip 10pt}
 
\author{Patrick Irwin\thanks{Email address: irwin@lps.umontreal.ca}\\
  {\normalsize {\sl Groupe des Physique des Particules, D\'{e}partement 
    de Physique,}}\\
  {\normalsize {\sl Universit\'{e} de Montr\'{e}al, C.P. 6128 succ. 
    Centre-Ville,}}\\
  {\normalsize {\sl Montr\'{e}al, Qu\'{e}bec, H3C\, 3J7, Canada.}}}

\maketitle
The energy radiated during the scattering of SU(3) monopoles is estimated 
as a function of their asymptotic velocity $v$. In a typical 
scattering process the total energy radiated is of order $v^3$ 
as opposed to $v^5$ for SU(2) monopoles. For charge 
$(1,1)$ monopoles the dipole radiation produced is estimated for 
all geodesics on the moduli space. For charge $(2,1)$ monopoles the
dipole radiation is estimated for the axially symmetric geodesic. The 
power radiated appears to diverge in the massless limit. The implications of 
this for the case of non-Abelian unbroken symmetry are discussed. 

\section{Introduction}
\news
\ \indent

The moduli space approximation has proved to be a very useful tool 
in the study of magnetic monopoles. It assumes that at low velocities the 
dynamics of BPS monopoles is determined by the geodesic motion in the 
manifold of static BPS configurations. This ignores the effect of 
Lorentz contraction on the monopole and a similar approximation is 
made regarding the total electric charge of the monopole. For a Lorentz 
boosted BPS monopole with rest mass $M$ and velocity $v$, the true 
kinetic energy is $\frac{1}{2}M\sqrt{1-v^2}$ whereas the moduli space 
approximation gives  $\frac{1}{2}Mv^2$. The approximation is thus 
accurate to order $v^4$, a similar statement is true regarding the 
total electric charge $q$ of the monopole.

However when more than one monopole is present another question needs to be 
considered. As the monopoles approach each other they will accelerate, and 
because these are charged objects, radiation will be produced. The 
amount of radiation measures the deviation from the moduli space 
approximation. In \cite{MS}, Manton and Samols estimate the radiation 
from the scattering of two SU(2) monopoles. To a first approximation 
the authors assume that the heavy inner cores of the monopoles 
evolve according to the moduli space approximation. 
The core region is where the monopole fields are non-Abelian and it is where
the energy density is concentrated. Outside this core region the 
fields are essentially Abelian.
The radiation of massless fields that such a motion 
of the monopoles cores would produce was then estimated. 
Corrections to this will arise because of deviations of the monopole 
core fields to the geodesic approximation. These corrections 
were shown to be of sub-leading order and can be ignored. 
The authors assume, as will be done here, that the 
dominant energy loss is to long range massless fields, as opposed to the 
short range massive fields. SU(2) monopoles have only one massless 
field; outside their core they can be treated using the usual
electrodynamics. To calculate the radiation one needs the 
multipole moments of the scalar, magnetic and electric fields 
which depend on the time dependent moduli space parameters. These give 
a field expansion of the monopole just outside the core region.
The radiation produced from this region to infinity can then be 
calculated using the usual formulae for multipole radiation from 
electrodynamics (see \cite{MS} for a more detailed description). 
For two SU(2) BPS monopoles their scalar and magnetic dipole 
moments always vanish. The leading 
order scalar and magnetic radiation is given by the quadrupole radiation. 
The quadrupole radiation in the head-on collision of two pure 
monopoles (each of mass 4$\pi$) was found to be approximately $17v^5$, 
where the asymptotic monopole velocity is $v$. This is likely to 
be the maximally radiating scattering geodesic due to the head-on nature 
of the collision. For dyons, the electric dipole moment can be nonzero, 
but with the restriction that the relative electric charge $q$ 
satisfies $q\approx v\ll 1$, the electric dipole 
radiation is also order $v^5$ \cite{MS}.

More generally, the total time where radiation is produced in a scattering 
event is of order $R/v$, $v$ is the incoming velocity and $R$ is the 
separation where inter-monopole forces become appreciable, which is 
roughly their core radius. The monopole core radius is of order $1/M$
where $M$ is the mass, so $R\approx 1/M$. Denoting the dipole moment 
by $d_i$ and the quadrupole moment by $Q_{ij}$, then an order of 
magnitude calculation gives
\be\label{1.1}
\frac{d^2}{dt^2}d_i\approx Mv^2\;,
\quad\qquad\frac{d^3}{dt^3}Q_{ij}\approx Mv^3\;.
\ee
The factors of $M$ are determined on dimensional grounds.
Inserting this into the formulae for dipole and quadrupole radiation
and multiplying by the time, $1/(Mv)$, gives an order of magnitude 
estimate of the dipole radiation in terms of the velocity as being of 
order $Mv^3$ and the quadrupole radiation of order $Mv^5$.
The energy radiated from magnetic or scalar quadrupoles will generally be 
of order $Mv^5$, and the result of Manton and Samols is very likely to hold
for higher charge SU(2) monopoles. Magnetic or scalar dipoles will 
radiate energy of order $v^3$ during a scattering 
process, but in the center of mass frame the scalar and magnetic dipole 
moments of all SU(2) monopoles vanish. It can similarly be shown 
that higher order multipole radiation is of higher order in $v$.

For the case of SU(3) monopoles with unbroken gauge group U(1)$\times$U(1) 
one might expect the same arguments used for the SU(2) case will also hold 
here since outside the monopole cores the theory reduces to that of  
U(1)$\times$U(1) gauge theory. However a difference arises in the 
long range behavior of the monopole fields, which in general have 
non-vanishing dipole moments. One can easily re-derive the formula for 
the energy radiated by a time dependent dipole in the U(1)$\times$U(1) 
case. The formula for the radiated energy is just a sum over the 
two U(1) factors and we find the result that the energy radiated 
in a typical scattering process is order $v^3$. As such, this 
does not pose any problems as regards the validity 
of the moduli space approximation, the kinetic energy is order $v^2$, 
so $E_{rad}/E_{kin}\approx v$, for small monopole velocities the radiation 
produced is small, providing the coefficient of $v^3$ in $E_{rad}$ is finite.

First we consider the case of charge $(1,1)$ monopoles. Their dipole 
moments are easily determined since the point description of these monopoles 
is valid. Assuming the motion of the monopoles inner core is that of a 
geodesic on the moduli space the total dipole radiation can be 
calculated to be $v^3$ times a function of two geometric parameters (which 
depend on the monopole geodesic) and the monopole masses.
The radiation produced remains small if one of the monopoles masses 
approaches zero. 

For $(2,1)$ monopoles, due to the complexity of the metric  
we only calculate the dipole radiation for a special axially
symmetric geodesic. Again the dipole radiation produced is a function times 
$v^3$. The case we are really interested in is the limit of non-Abelian 
unbroken symmetry which occurs when the mass of the $(0,1)$ monopole is taken 
to zero. Taking the naive limit, the function multiplying $v^3$ diverges 
implying that infinite radiation is produced and thus invalidating the 
moduli space approximation. The argument given around Eq. (\ref{1.1}) for 
SU(2) monopoles does not work here as now there are two mass parameters, 
it is a function of their ratio that diverges.

However since the monopoles are finite energy objects, it is not possible 
for them to produce infinite radiation, therefore there is a problem with 
our naive reasoning. This
can be seen by examining the core size of the monopole in the massless 
limit. For a monopole of mass $m$ its core size is of order $1/m$, 
this diverges in the massless limit. As the mass $m$ of the $(0,1)$ 
monopole approaches zero, the size of its core region expands; 
for $m$ small enough the core of the $(0,1)$ monopole will
generically surround the cores of the two massive monopoles, \cite{LWY1}. 
Inside this overall core region the fields are non-Abelian. However the 
energy density is concentrated in a small region around the $(2,0)$
monopole. Just outside the overall core region, 
where we measure the multipole moments, the multipole
expansion will differ significantly from that of a static configuration. 
This is because the fields in the core region become highly relativistic 
in the massless limit. The assumption that the monopole fields inside 
the core region are well approximated by the geodesic approximation  
breaks down in the massless limit. Our results are therefore 
somewhat inconclusive, nonetheless in the massless limit it appears that 
the time dependent monopole fields do have large deviations from that 
predicted by the moduli space approximation. 

In \cite{MS}, and in the present work, the amount of radiation produced 
is used as a estimate of the validity of the moduli space 
approximation. In \cite{Stuart}, Stuart proves the validity of the geodesic 
approximation at low velocities for SU(2) monopoles 
using rigorous analytical methods. It would be interesting if the methods in 
\cite{Stuart} could be extended to higher gauge groups, indeed it seems that 
such an approach is necessary as the calculations presented here do not
have sufficient rigor to confirm or reject the validity of the moduli 
space approximation in the massless limit.

\section{The radiation formula for SU(3) monopoles}
\news
\ \indent

We assume that the Higgs
field is in the adjoint representation and we are in the BPS limit in
which the scalar potential is zero but a nonzero Higgs expectation
value is imposed at spatial infinity as a boundary condition.
An SU(3) gauge theory can be broken by an adjoint Higgs mechanism
to either U(1)$\times$U(1) or U(2). The generators of SU(3) may be 
chosen to be two commuting operators $H_i$, $i=1, 2$,
with $\mbox{Tr}\,H_iH_j=\delta_{ij}$,
together with ladder operators associated with the roots 
$\pm\alphabf$, $\pm\betabf$, $\pm(\alphabf+\betabf)$  that obey 
\be
[H_i,\mbox{E}_{\gamma}]=\gamma^i\mbox{E}_{\gamma},\;\;\;\;\; 
[\mbox{E}_{\gamma},
\mbox{E}_{-\gamma}]=\gammabf\cdot\bf{H}=H_{\gamma}\;,
\ee
for $\gammabf$ any root. Define 
$\alphabf=(-1/2,\sqrt{3}/2)$ and $\betabf=(1,0)$.
We choose the singular gauge where the Higgs field $\phi$ is constant at
spatial infinity, equal to\, $\phi_{\infty}$.
Choosing this constant value to lie
in the Cartan sub-algebra defines a vector $\bf{h}$ by   
$\phi_{\infty}=\bf{h}\cdot\bf{H}$.
If SU(3) is broken to U(1)$\times$U(1), all roots have nonzero inner
product with $\bf{h}$ and there is a unique set of simple roots with
positive inner product with $\bf{h}$. If SU(3) is broken to U(2)
then one of the roots, $\betabf$ say, is perpendicular to $\bf{h}$. 

For any finite energy solution, asymptotically 
\be
B_i=\frac{x_i}{4\pi|{\bf x}|^3}G\,,
\ee
$G$ is a constant element of the Lie algebra of SU(3). The Cartan sub-algebra 
may be chosen so that $G=\bf{g}\cdot\bf{H}$. 
The quantization of magnetic charge determines $\bf{g}$ as 
\be\label{2.3}
{\bf{g}}=\frac{4\pi}{e}\left\{k_{\alpha}\alphabf+k_{\beta}
\betabf\right\}\;,\qquad
\ee
where $e$ is the gauge coupling, and $k_{\alpha}$, 
$k_{\beta}$ are non-negative integers.
Such a solution is called a $(k_{\alpha},k_{\beta})$ monopole and has mass 
${\bf g}\cdot{\bf h}$.

When SU(3) is broken to U(1)$\times$U(1) the topological
charges of the monopoles are determined by the integers 
$k_{\alpha}$, $k_{\beta}$.
All BPS monopoles may be thought of as superpositions of two 
fundamental monopoles given by embeddings of the charge one SU(2) 
monopole \cite{Wein1}. Associated with each root is an SU(2) sub-algebra.
The two fundamental monopoles are obtained by embedding the charge one
SU(2) monopole along the SU(2) sub-algebra associated to the  
simple roots, $\alphabf$ and $\betabf$ \cite{Wein1}. 
Each fundamental monopole has four
zero modes, corresponding to its position and a U(1) phase. Embedding along 
the root $\alphabf$ gives the $(1,0)$ (or $\alphabf$) monopole
charged with respect to one of the unbroken U(1)'s. Similarly, one can 
embed along the root $\betabf$ to give the $(0,1)$ (or $\betabf$)
monopole charged with respect to the other unbroken U(1) group. 
A $(k_{\alpha},0)$ monopole or a
$(0,k_{\beta})$ monopole is made up of only one type of monopole 
and behaves like the corresponding SU(2) monopole. A 
$(k_{\alpha},k_{\beta})$ monopole has mass 
$(k_{\alpha}M+k_{\beta}m)$, with $M$ is the mass of a $(1,0)$ monopole
and $m$ the mass of a $(0,1)$ monopole. 
BPS monopoles interact in different ways depending on whether they
are of the same type or of different types. The metric on the moduli
space of two SU(3) monopoles of different type, the $(1,1)$ monopole
is the Taub-NUT metric. The metric for a $(2,0)$ monopole 
is the Atiyah-Hitchin metric. The metric on the moduli space of 
$(2,1)$ monopoles mixes these interaction types.


We now consider the multipole radiation.
For SU(2) monopoles a core region exists where the fields are 
non-Abelian. Outside this core region the fields 
can be gauge transformed to be exponentially 
close to Abelian fields, i.e. proportional to the same SU(2) generator. 
The core region is where the energy density of a monopole is concentrated.
Outside this region the fields satisfy the source-less Maxwell equations.
As regards the asymptotic region, the fields in the core region
can be viewed as providing effective U(1) sources for the asymptotic 
fields. Just outside the core region the monopole fields are Abelian, and 
assuming no appreciable radiation has been emitted by the slow moving core, 
the fields are determined from the moduli space approximation. For example,
the scalar field can be expanded as  
\be
\phi=\tau_3\{v-\frac{g}{4\pi|{\bf x}|}+
\frac{{\bf d}_s(t)\cdot\hat{{\bf x}}}{4\pi |{\bf x}|^2}+\dots\}\quad,
\ee
where ${\bf d}_s(t)$ is the time dependent dipole moment, its time dependence
is determined from the moduli space approximation, and $\tau_3$ is a fixed 
element of the SU(2) Lie algebra. Because outside of the core 
the monopole behavior is exactly that of U(1) dyons, the multipole 
radiation formulae is unchanged from the U(1) theory.

Turning now to the case of SU(3) broken to U(1)$\times$U(1),
almost everything said for SU(2) monopoles carries over here. Again there 
exists a core region outside of which the fields can be gauge transformed 
to lie in the Cartan sub-algebra,
the fields satisfy the source-less U(1)$\times$U(1) Maxwell wave 
equation outside the core. The core region again provides effective 
U(1)$\times$U(1) sources for the asymptotic region. Using the orthogonal 
basis $H_1$, $H_2$, the field equations decompose into two linearly 
independent parts and the radiation formula can be derived as
before. Each multipole moment has a $H_1$ and a
$H_2$ component. Just outside the core region the scalar field 
may be expanded as
\be
\phi={\bf h}\cdot{\bf H}-\frac{{\bf g}\cdot{\bf H}}{4\pi|{\bf x}|}+
\frac{{\bf d}_s(t)\cdot\hat{{\bf x}}}{4\pi|{\bf x}|^2}+\dots\quad,
\ee
where ${\bf d}_s(t)={\bf d}^1_s(t)\,H_1+{\bf d}^2_s(t)\,H_2$,      
is the time dependent scalar dipole moment.
The radiation formula is just a sum of the radiation produced 
from each component. Scalar, magnetic and electric dipole radiation will be 
produced. The general dipole radiation 
formula for SU(3) monopoles with maximal symmetry breaking is
\be\label{power}
P(t)=\frac{1}{6\pi}\sum_{i=1,2}
[\ddot{{\bf d}}^i_s\cdot\ddot{{\bf d}}^i_s+
\ddot{{\bf d}}^i_b\cdot\ddot{{\bf d}}^i_b+\ddot{{\bf d}}^i_e
\cdot\ddot{{\bf d}}^i_e]\;,
\ee
where the subscripts $s$, $b$, $e$ denote the scalar, magnetic and 
electric dipole moments respectively.
The total power radiated, $P$, is just the time integral of $P(t)$,
$P=\int^{\infty}_{-\infty}dt\,P(t)$. We now consider in turn the 
case of $(1,1)$ and $(2,1)$ monopoles. The procedure
is to find the dipole moments as a function of the moduli space parameters. 
The time dependence of the moduli space parameters is determined from the 
geodesic equation. The radiated power from Eq. (\ref{power}) is then 
integrated over the corresponding geodesic in the moduli space.

\section{Charge $(1,1)$ monopoles}
\news
\ \indent

Charge $(1,1)$ monopoles are relatively easy to describe due to their simple
point-like interactions. The monopole consists of two embedded charge one 
SU(2) monopoles, embedded along distinct roots, $\alphabf$ and $\betabf$.
We want an expression for the dipole moments of the monopole fields, 
it turns out that these are obtained from a simple addition of the dipole
moments of each embedded monopole. A charge one SU(2) monopole with electric 
charge $q$, magnetic charge $g$, positioned at ${\bf X}\in \R^3$,
may be written in the singular gauge as 
\be
\phi({\bf x}) = \frac{(q^2+g^2)^{1/2}}{4\pi|{\bf x}-{\bf X}|}\tau_3\;,\;\;
B_i({\bf x}) = \frac{g({\bf x}-{\bf X})}{4\pi|{\bf x}-{\bf X}|^3}\tau_3\;,\;\;
E_i({\bf x}) = \frac{q({\bf x}-{\bf X})}{4\pi|{\bf x}-{\bf X}|^3}\tau_3\;.
\ee
This expression for the fields is valid outside the monopole core with 
exponentially small corrections, and is just an embedding of a U(1) dyon 
into SU(2). For charge $(1,1)$ SU(3) monopoles a similar expression can 
be written down for the fields. In \cite{WY}, the Nahm transform 
was inverted for a variety of monopole charges to give explicit 
expressions for the monopole fields. For $(1,1)$ SU(3) monopoles, the 
long range fields are just a naive 
sum of the embedded SU(2) fields. Inside the monopoles cores and near 
the line joining the monopoles the expressions for the fields are 
more complicated. The fields considered in \cite{WY} were time independent, 
i.e. the moduli space parameters were time independent. 
Electric fields are absent in this case. Introducing time dependence of the 
moduli parameters generally gives rise to electric fields. The results in 
\cite{WY} show that when the electric fields are zero the multipole moments 
are just a naive sum of those from the individual monopoles. We assume that 
this is also true for non-zero electric fields. We can then write down the 
long range $(1,1)$ monopole fields. They are just what one obtains from an 
ansatz of point electric, magnetic and scalar charges .
Using the root basis described earlier, a $(1,1)$ monopole  
with composed of a $(1,0)$ monopole with electric charge $q_1$ and a 
$(0,1)$ monopole with electric charge $q_2$ can be asymptotically written as
\bea
\phi({\bf x})&=&{\bf h}\cdot{\bf H}
+ \frac{(q_1^2+g^2)^{1/2}}{4\pi|{\bf x}-{\bf X}_1|}H_{\alpha}+
\frac{(q_2^2+g^2)^{1/2}}{4\pi|{\bf x}-{\bf X}_2|}H_{\beta}\;,\\
B_i({\bf x})&=&\frac{g({\bf x}-{\bf X}_1)}{4\pi|{\bf x}-{\bf X}_1|^3}
H_{\alpha}+\frac{g({\bf x}-{\bf X}_2)}{4\pi|{\bf x}-{\bf X}_2|^3}H_{\beta}
\;,\nonumber\\
E_i({\bf x})&=&\frac{q_1({\bf x}-{\bf X}_1)}{4\pi|{\bf x}-{\bf X}_1|^3}
H_{\alpha}+\frac{q_2({\bf x}-{\bf X}_2)}{4\pi|{\bf x}-{\bf X}_2|^3}
H_{\beta} \;.\nonumber
\eea
Here ${\bf X}_1$ and ${\bf X}_2$ are the positions of the two 
monopoles in space. We can read off the dipole moments which are
given by 
\bea
{\bf d}_s&=&(q_1^2+g^2)^{1/2}\,{\bf X}_1\,H_{\alpha}+
(q_2^2+g^2)^{1/2}\,{\bf X}_2\,H_{\beta}\\
{\bf d}_b&=&-(g\,{\bf X}_1\,H_{\alpha}+g\,{\bf X}_2\,H_{\beta})\nonumber\\
{\bf d}_e&=&-(q_1\,{\bf X}_1\,H_{\alpha}+q_2\,{\bf X}_2\,
H_{\beta})\;.\nonumber
\eea
To use the formula in Eq. (\ref{power}) 
for the power radiated by scalar, electric and 
magnetic dipoles we have to decompose the dipole moments into orthogonal 
parts of the Lie algebra. Using our previous definitions for the roots 
$\alphabf$, and $\betabf$ we have $H_{\alpha}=\frac{1}{2}(-H_1+\sqrt{3}H_2)$
and $H_{\beta}=H_1$.
We then decompose the dipole 
moments into their $H_1$ and $H_2$ components, ${\bf d}={\bf d}_1\,H_1+
{\bf d}_2\,H_2$ and using Eq. ({\ref{power}) we get
\bea
P(t)&=&\frac{1}{6\pi}\{2(g^2+q_1^2)\ddot{{\bf X}}_1\cdot\ddot{{\bf X}}_1+
2(g^2+q_2^2)\ddot{{\bf X}}_2\cdot\ddot{{\bf X}}_2\\
&-&[g^2+q_1q_2+(g^2+q_1^2)^{1/2}(g^2+q_2^2)^{1/2}]
\ddot{{\bf X}}_1\cdot\ddot{{\bf X}}_2\} 
\;.\nonumber
\eea

Our assumption that the monopole cores evolve according 
to the moduli space approximation implies that the monopoles
centers ${\bf X}_1$, ${\bf X}_2$ are determined by a geodesic 
on the moduli space. The overall center of mass, 
${\bf R}=(M{\bf X}_1+m{\bf X}_2)/(M+m)$, can be ignored as it evolves 
with constant velocity. The relative moduli space has the Taub-NUT 
metric with positive mass \cite{Connell, GL}. This is a four dimensional 
manifold whose parameters describe the relative position, ${\bf r}$, 
and phase, $\chi$,
of the monopole. The geodesics on the moduli space manifold are equivalent
to the equations of motion derived from a Lagrangian. This Lagrangian
can be derived from the point-particle interactions of the monopoles. 
The Lagrangian is given by
\be\label{metric}
L=\frac{1}{2}(\mu+\frac{g^2}{8\pi r})\dot{{\bf r}}\cdot\dot{{\bf r}}
+\frac{1}{2}
(\frac{g^2}{8\pi})^2(\mu+\frac{g^2}{8\pi r})^{-1}(\dot{\chi}+
{\bf w}({\bf r})\cdot\dot{\bf r})^2 \;,
\ee
with $\mu$ equal to the reduced mass, $\mu=Mm/(M+m)$. 
The relative monopole position, ${\bf r}={\bf X}_1-{\bf X}_2$, 
is written in spherical polar coordinates $r$, $\theta$, $\phi$, and 
${\bf w}({\bf r})$ is the Dirac monopole potential satisfying, 
$\nabla\times{\bf w}({\bf r})={\bf r}/r^3$.
The conserved relative electric charge of the monopole, $q$, is conjugate
to $\chi$, i.e. 
$q=(g^2/8\pi)^2(\mu+g^2/8\pi r)
^{-1}(\dot{\chi}+{\bf w}({\bf r})\cdot\dot{\bf r})\,$.
In terms of the monopoles individual charges $q_1$ and $q_2$, the 
relative electric charge $q$ is given by $q=(q_1-q_2)/2$. There 
also exists the conserved total electric charge $Q$, with
\be
Q=\frac{Mq_1+mq_2}{M+m}\;.
\ee
We henceforth set $Q=0$ to restrict to motion on the relative moduli space. 
This determines $q_1$ and $q_2$ as functions of the relative electric charge 
$q$,
\be\label{q}
q_1=\frac{2mq}{M+m}\;,\qquad q_2=-\frac{2Mq}{M+m}\;.
\ee
${\bf X}_1$, ${\bf X}_2$ are given in terms of ${\bf r}$ and ${\bf R}$ by
\be
{\bf X}_1={\bf R}+\frac{m}{M+m} {\bf r}\;,\qquad
{\bf X}_2={\bf R}-\frac{M}{M+m}{\bf r}\;.
\ee
Using this in the above formula for the power radiated we have
\be\label{2.6}
P(t)=\lambda\,\ddot{{\bf r}}\cdot\ddot{{\bf r}}\;,
\ee
\bea\label{2.6'}
\lambda&=&\frac{1}{3\pi}\{\frac{m^2(g^2+q_1^2)+(g^2+q_2^2)M^2}{(M+m)^2}\}\\
&+&\{\frac{mM}{6\pi(M+m)^2}(g^2+q_1q_2+
(g^2+q_1^2)^{1/2}(g^2+q_2^2)^{1/2})\}\;.\nonumber
\eea
This formula contains terms of different order in the velocity $v$, 
which is the order parameter in the problem. The electric charges 
$q_1$ and $q_2$  are really small parameters of the same 
order as the velocity in the moduli space approximation. So the terms 
involving the $q_1,$ $q_2$ have the same order of magnitude as the 
quadrupole terms arising from the scalar and magnetic radiation. 
We retain the $q_1,$ $q_2$ terms just to give a complete formula 
for the dipole radiation. It now remains to calculate 
$\int\ddot{{\bf r}}\cdot\ddot{{\bf r}}$
for a given geodesic on the moduli space. 
We fix the coupling constant by setting $g=4\pi$.
The equation of motion for ${\bf r}$ resulting from the Lagrangian, 
Eq. (\ref{metric}),
is
\be
(\mu+\frac{2\pi}{r})\ddot{{\bf r}}= -\frac{2\pi}{r^3}
\{\frac{1}{2}(\dot{\bf r}\cdot\dot{\bf r}){\bf r}-
(r\dot{r})\dot{\bf r}\}+q\frac{\dot{\bf r}\times{\bf r}}
{r^3}+\frac{q^2}{4\pi r^3}{\bf r}\;.
\ee
From this it is not too hard to show that 
\be\label{rr}
\ddot{{\bf r}}\cdot\ddot{{\bf r}}=\frac{4\pi^2E}{(\mu r^4+2\pi)^4}\;,
\ee
with $E$ is the conserved energy given by 
\be\label{energy}
E=\frac{1}{2}(\mu+\frac{2\pi}{r})\{\dot{{\bf r}}\cdot\dot{{\bf r}}  
+(\frac{q}{2\pi})^2\}\;.
\ee
Equation (\ref{rr}) simplifies considerably the task of calculating the power 
radiated as now all that is needed is to determine the separation parameter 
$r$ as a function of time.
So the total power radiated is
\be
P=4\pi^2E^2\lambda\int_{-\infty}^{\infty}dt\frac{1}{(\mu r +2\pi)^4}\;. 
\ee
We now change the integration variable from $t$ to $r$ to yield
\be\label{powerr}
P=8\pi^2E^2\lambda\int_{r_{min}}^{\infty}dr\frac{1}
{\dot{r}(\mu r +2\pi)^4}\;. 
\ee
This is valid since geodesics on the Taub-NUT space are hyperbolae 
\cite{Connell}, so in a scattering process 
$r$ asymptotically approaches infinity and there is only one turning point.
We need to determine $\dot{r}$ as a function of $r$, in order to 
do this we examine the equations of motion and use 
the conserved quantities. In addition to the energy there are 
two conserved vector quantities, the angular momentum ${\bf J}$, 
and the vector ${\bf K}$. The existence of the conserved vector 
quantity ${\bf K}$ is a special feature of the Taub-NUT manifold,
\cite{GM}, owing its existence to the self dual nature of the metric.
Defining ${\bf p}$ as
\be
{\bf p}=(\mu+\frac{2\pi}{r})\dot{{\bf r}}\;,
\ee 
${\bf J}$ and  ${\bf K}$ are given by
\bea\label{JK}
{\bf J}&=&{\bf r}\times{\bf p}-q\hat{{\bf r}}\;,\\
{\bf K}&=&{\bf p}\times{\bf J}-(2\pi E-\frac{\mu q^2}{2\pi})
\hat{{\bf r}}\;.\nonumber
\eea
It is not difficult to check from Eq. (\ref{metric})
that these vectors are indeed conserved. The angular momentum 
is a sum of the orbital angular momentum and the Poincar\'{e} 
contribution. The magnitude of the orbital angular momentum, 
$l=|{\bf r}\times{\bf p}|$, is also conserved and satisfies
\be
J^2=l^2+q^2\;,
\ee
with $J^2={\bf J}\cdot{\bf J}$.
We now choose a coordinate frame so that the $\hat{z}$ component of ${\bf r}$ 
is constant. This is achieved by taking the vector ${\bf K}-\kappa{\bf J}$ 
to point in the $\hat{z}$ direction ($\kappa$ is a constant defined below). 
Then writing 
\bea\label{318}
{\bf K}-\kappa{\bf J}&=&\frac{J^2-q^2}{z_0}\hat{z}\;,\\
\kappa&=&\frac{1}{q}(2\pi E-\frac{\mu q^2}{2\pi})\nonumber\;,
\eea
implies the $\hat{z}$ component of ${\bf r}$ is constant, equal 
to $z_0$. This can be checked using Eq. (\ref{JK}). 
The separation vector is then written in cylindrical coordinates 
($\rho$, $\psi$, $z$) as 
\bea
{\bf r}&=&\rho\,\hat{p}+z_0\,\hat{z}\;,\\
\dot{{\bf r}}&=&\dot{\rho}\,\hat{\rho}+\rho\,
\dot{\psi}\,\hat{\psi}\;.\nonumber
\eea
The relative monopole motion is thus in a fixed plane. 
The angular momentum ${\bf J}$ may be calculated in this basis,
${\bf J}=J_{\rho}\,\hat{\rho}+J_{\psi}\,\hat{\psi}+J_{z}\,\hat{z}$, with
\be\label{J}
J_{\rho}=-\frac{\rho q}{r}-(\mu+\frac{2\pi}{r})\rho\dot{\psi}z_0\;,\quad
J_{\psi}=(\mu+\frac{2\pi}{r})\dot{\rho}z_0\;,\quad
J_z=(\mu+\frac{2\pi}{r})\rho^2\dot{\psi}-\frac{qz_0}{r}\;,
\ee
and $r=(\rho^2+z_0^2)^{1/2}$. From the equation for $J_z$ in (\ref{J}) 
we can express $\dot{\psi}$ in terms of $\rho$. We then use the energy
equation (\ref{energy}) to determine $\dot{\rho}$.
Using $\dot{{\bf r}}\cdot\dot{{\bf r}}=\dot{\rho}^2+\rho^2\,\dot{\psi}^2$,
Eq. (\ref{energy}) may be converted into an equation for $\dot{\rho}$ and 
since $r\dot{r}=\rho\dot{\rho}$, this can then be transformed 
into an equation for $\dot{r}$ in terms of $r$.
Finally we need to determine the constants $J_z$ and $z_0$.
The constant $J_z$ can be seen to be $J_z=-\kappa z_0$ using
\be
\label{Jz}J_z={\bf J}\cdot\frac{({\bf K}-\kappa{\bf J})}
{|{\bf K}-\kappa{\bf J}|}\;.
\ee
Then using the above expressions for ${\bf J}$ and ${\bf p}$ in cylindrical 
coordinates we evaluate $({\bf p}\times{\bf J})_3$ and insert into
Eq. (\ref{318}) using the definition of ${\bf K}$. Comparing both sides of 
the first equation in (\ref{318}) we get
\be\label{z0}
z_0=\frac{q\{J^2-q^2\}^{1/2}}{2\pi E}\;.
\ee  
If $z_0=0$, both 
monopoles move in the same plane. This occurs if $q=0$ or $J^2=q^2$ ($l=0$), 
in the latter case the scattering is along a straight line. 
It can be checked that Eq. (\ref{318}) has a sensible limit 
as $z_0\rightarrow 0$. Using these values for $J_z$ and $z_0$ 
we get the following equation for $\dot{r}$,
\be
\dot{r}= \frac{1}{(\mu r+2\pi)}\{
2\mu r^2(E-\frac{\mu q^2}{8\pi^2})+4\pi r(E-\frac{\mu q^2}{4\pi^2})
-J^2\}^{1/2}\;.
\ee
Recalling Eq. (\ref{powerr}), the limits of the integral 
are $r_{min}$ and $\infty$,
$r_{min}$ is found by solving $\dot{r}=0$. The energy $E$ can be 
expressed in terms of the asymptotic velocity, $v$, using Eq. (\ref{energy})
the asymptotic velocity is determined from 
\be\label{E}
E=\frac{\mu}{2}\{v^2+(\frac{q}{2\pi})^2\}\;.
\ee 
The magnitude of the conserved orbital angular momentum $l$ 
($l^2=J^2-q^2$) can be
written as $l=\mu vr_0$. Here $r_0$ is the asymptotic impact parameter.
$r_0$ satisfies $r_0\geq z_0$ because the monopoles separation is 
constant in the $\hat{z}$ direction, their relative velocity is in 
the $x$-$y$ plane. $r_0^2=z_0^2+r_{\perp}^2$ where $r_{\perp}$ is the 
planar impact parameter. The resulting integral for the radiated 
power can be done without too much difficulty with the result 
\be\label{result}
P=\frac{\lambda v^3}{r_0}g(y)\;,
\ee
where 
\be
g(y)=\frac{1}{y^2}\{[1+\frac{3}{y^2}]
[\frac{\pi}{2}-\sin^{-1}(1+y^2)^{-1/2}]-\frac{3}{y}\}\;,\qquad
y=\frac{\mu r_0}{\pi(1+q^2/4\pi^2v^2)}\;.
\ee
This is the result for the dipole radiation produced during a 
scattering of $(1,1)$ monopoles. The result depends on the incoming velocity 
$v$, the relative electric charge $q$, the impact parameter $r_0$
and the monopole masses $M$, $m$.
Recall from Eq. (\ref{2.6'}) and Eq. (\ref{q}) that
$\lambda$ is dependent on $q$, $M$ and $m$. 
The function $g(y)$ satisfies 
\be
\quad \lim_{y\rightarrow 0}g(y)=\frac{4y}{15}\;,
\quad\lim_{y\rightarrow\infty}g(y)=\frac{\pi}{2y^2}\;.
\ee
For positive $y$, $g(y)$ has one turning point, its maximum, at 
$y\approx 1.1$, where $g(y)\approx 0.15$.
The power radiated is maximal for the head on 
collision of two pure monopoles as might be expected, the case 
of $r_0=q=0$ with $P=4\,\lambda\,\mu\,v^3/15\pi$. In terms of $v$, the  
relative velocity, the power radiated has leading order $v^3$.
If one of the monopole masses approaches zero ($\mu\rightarrow 0$), the 
total radiation, $P$, correspondingly decreases to zero.
For large values of the impact impact parameter $r_0$, 
$P$ falls off as roughly as $v^3/r_0^3$.

\section{Charge $(2,1)$ monopoles}
\news
\ \indent

For the previous case of charge $(1,1)$ monopoles 
we were able to find a explicit formula for the dipole
radiation produced from any geodesic motion. For charge $(2,1)$ 
monopoles we can also compute the dipole moments and the moduli space 
metric is known. But the calculations are much more involved given the 
complexity of the $(2,1)$ metric. We will restrict ourselves to computing 
the radiation produced from a single axially symmetric 
geodesic where the calculations simplify considerably.
This scattering event was described in \cite{HIM}, it involves a 
head-on collision of the monopoles, we are especially interested in 
the massless limit. It is very likely that the radiation produced 
in this scattering event will be the maximal of all scattering events,
again due to the head-on nature of the collision. 
The procedure of finding the radiation produced is exactly the same as 
the previous case; first find the dipole moments as a time dependent 
function of the moduli space parameter, the explicit time dependence is 
determined from the equation for geodesic motion. This is inserted into the
dipole radiation formula and integrated over time to give the total
power radiated.

The dipole moment of a $(2,1)$ monopole configuration is not hard to 
evaluate. A $(2,1)$ monopole can generally be thought of as a $(2,0)$ 
monopole (an embedded charge two SU(2) monopole of mass $2M$) combined with 
a $(0,1)$ monopole (an embedded charge one SU(2) monopole of mass $m$), in 
much the same way as a $(1,1)$ monopole is a combination 
of a $(1,0)$ monopole with a $(0,1)$ monopole. We choose the coordinate 
system where the $(2,0)$ monopole center of mass is at the origin. 
When the $(2,0)$ monopole is centered it has no dipole moment.
The total dipole moment is deduced from that of the $(0,1)$ monopole. 
The $(0,1)$ monopole has a well defined 
position and the point approximation can be used as before to calculate
its dipole moment. The above statements can be proved in the case of the 
axially symmetric geodesic. To show this, first notice that the dipole 
moment must point along the axis of symmetry. In the singular gauge 
introduced earlier we can write the Higgs field as
\be
\phi=v\,{\bf h}\cdot{\bf H}-\frac{(2\alphabf+\betabf)
\cdot{\bf H}}{4\pi|{\bf x}|}+\frac{({\bf d}^
{\alpha}H^{\alpha}+{\bf d}^{\beta}H^{\beta})\cdot\hat{{\bf x}}}
{4\pi|{\bf x}|^2}+...\quad.
\ee
The fields are invariant up to gauge transform under rotations about the 
$\hat{z}$ axis. To leave invariant $v\,{\bf h}\cdot{\bf H}$ and 
$(2\alphabf+\betabf)\cdot{\bf H}/|{\bf x}|$, 
the compensating gauge transform must have 
its constant and $1/|{\bf x}|$ terms in the Cartan sub-algebra ${\bf H}$. 
Therefore the $1/|{\bf x}|^2$ component of $\phi$ is also 
unchanged by the gauge 
transform. This implies that the $1/|{\bf x}|^2$ component of $\phi$ 
is invariant under a rotation about the $\hat{z}$ axis without a 
corresponding gauge transform, which in turn means that 
$d^{\alpha}_x=d_x^{\beta}=d^{\alpha}_y=d_y^{\beta}=0$. 
So the dipole moment points in the $\hat{z}$ direction and can be 
calculated knowing just the fields along the axis of symmetry which we 
compute below from the Nahm transform. This argument does not hold in the 
massless limit where the unbroken gauge symmetry is enhanced. Indeed the 
dipole moments of the fields have been analyzed for the minimal symmetry 
breaking case, \cite{Sch}, and such a point interpretation is not possible.

The Nahm data is known explicitly for the axially symmetric 
configurations. Using the Nahm transform we can invert this 
to derive the monopole fields. To derive the monopole fields at a given 
point in space it is necessary to solve a linear equation with the Nahm 
data and the aforementioned point as input, from the solution
to this equation one  can find the monopole fields. It turns out that 
this equation is difficult to solve for points off the axis of symmetry, but
the solution can be found for points on the axis of symmetry. 
Similar calculations have been done previously \cite{Dan, WY}, in the 
present case the calculations are long and not very illuminating, 
we will just state here the results. The electric fields are zero 
and the scalar dipole moment ${\bf d}_s$ is equal to the magnetic dipole 
moment ${\bf d}_b$. In the frame where the $(2,0)$ 
monopole is centered, the dipole moments are unchanged from those found 
in \cite{Dan} for $m=0$ (the higher multipole moments do depend on $m$).
As mentioned earlier the dipole moments for $m=0$ are non-zero off
the axis of symmetry, this is not the case here, where $m>0$.  
The Nahm transform can be used thus to determine the dipole moments on the 
axis of symmetry. Denoting ${\bf r}_{\beta}$ the position of the 
$(0,1)$ monopole, the dipole moments are given by
\be
{\bf d}_s={\bf d}_b=\{{\betabf}\cdot{\bf H}\}{\bf r}_{\beta}\;.
\ee
This is exactly what one derives assuming that the total dipole moment 
is determined from the point-like $(0,1)$ monopole. 
Inserting this in Eq. (\ref{power}) we get
the power radiated as 
\be\label{power21}
P=\frac{1}{3\pi}\int^{\infty}_{-\infty}dt\,\ddot{{\bf r}}_{\beta}\cdot
\ddot{{\bf r}}_{\beta}\;.
\ee
In fact it is not justified to restrict to the frame where the $(2,0)$ 
monopole is centered as it is not a geodesic sub-manifold of the full 
moduli space. We really should work in the overall center of mass frame. 
Returning to the overall center of mass frame has the effect of changing 
the mass parameter $m$ of the $(0,1)$ monopole to the reduced mass
in the metric, \cite{HIM}, in addition the formula for the power, 
(\ref{power21}), is multiplied by a function of $m/M$.
We will continue to use the $(2,0)$ monopole centered frame for 
simplicity as the formulas are more transparent.

In the notation introduced earlier we have two $\alphabf$ monopoles,
the $(2,0)$ monopole, and one $\betabf$ monopole, the $(0,1)$ monopole.
The axially symmetric geodesic that we consider has at one asymptote 
the spherically symmetric $\alphabf+\betabf$ monopole approaching the second 
$\alphabf$ monopole. The monopoles collide at the origin, the $\betabf$ 
monopole then scatters to spatial infinity, the two $\alphabf$ monopoles 
remain coincident. The configuration asymptotically resembles the $\betabf$ 
monopole separating from the charge two donut $\alphabf$ 
monopole. Axial symmetry is preserved at all times, see \cite{HIM} for 
more details.

The moduli space metric for charge $(2,1)$ monopoles is derived in 
\cite{HIM}. This metric is equivalent to a Lagrangian describing the 
monopole dynamics. We consider the axially symmetric monopoles which 
form a one dimensional geodesic sub-manifold of the moduli space.
The sub-manifold can be thought of as a union of two different regions.
The Lagrangian in Region 1 is given 
by $L=\frac{1}{2}a_1(D)\,\dot{D}^2$ or,
\bea\label{hypmetric}
L&=&\frac{1}{2}\{(\sinh DM\cosh DM-DM)(DM-\tanh DM)\frac{\cosh
DM}{2D\sinh^3 DM}\\
&+&\frac{m(\sinh DM\,\cosh DM -DM)^2}{4\sinh^4 DM}\}
\dot{D}^2\;.\nonumber
\eea
This region describes the $\alphabf$ monopole 
approaching and colliding with the spherically symmetric 
$\alphabf+\betabf$ monopole. The position 
of the $\betabf$ monopole is 
well defined and is given by 
\be\label{rhyp}
{\bf r}_{\beta}=(0,0,-\frac{D}{2}\coth{DM} )\;.
\ee
In this region $D$ satisfies $0\leq D <\infty$. The separation 
of the $\alphabf$ monopole from the $\alphabf+\betabf$ monopole 
is approximatively $D$ for large $D$. The collision occurs at $D=0$.
After the collision the two $\alphabf$ monopoles coalesce to form an
axially symmetric configuration and the $\betabf$ monopole
scatters to spatial infinity, this is Region 2. 
The Lagrangian is given in Region 2 by $L=\frac{1}{2}a_2(D)\,\dot{D}^2$  or,
\bea\label{trigmetric}
L&=&\frac{1}{2}\{(\sin DM\cos DM-DM)(DM-\tan DM)\frac{\cos
DM}{2D\sin^3 DM}\\
&+&\frac{m(\sin DM\,\cos DM -DM)^2}{4\sin^4 DM}\}
\dot{D}^2\;.\nonumber
\eea
This region describes the $\betabf$ monopole separating from the donut 
configuration of the two $\alphabf$ monopoles. The position of the 
$\betabf$ monopole is 
\be\label{rtrig}
{\bf r}_{\beta}=(0,0,-\frac{D}{2}\cot{DM})\;. 
\ee
The donut configuration is at the origin. $D$ is constrained to satisfy
$0\leq D<\pi/M$. As $D$ approaches $\pi/M$, the $\betabf$ monopole 
approaches spatial infinity.
The two regions fit together smoothly at $D=0$ and together form a
geodesic sub-manifold of the full moduli space. In fact there is a two 
dimensional family of axially symmetric geodesics, one can act on the 
above family with a U(1) factor conserving the axial symmetry. 
For simplicity we restrict here to the one dimensional case.

To determine the total power radiated we need to find 
$\ddot{{\bf r}}_{\beta}\cdot\ddot{{\bf r}}_{\beta}$, 
from Eq. (\ref{rhyp}) and Eq. (\ref{rtrig}) this 
can be expressed in terms of $D$ and its times derivatives.
$D$ is then determined as a function of time from the Lagrangian, 
Eq. (\ref{hypmetric}) and  Eq. (\ref{trigmetric}). We can express the total 
power radiated as that of a sum coming from Regions $1$ and $2$. 
Define the $\hat{z}$ component of ${\bf r}_{\beta}$ 
in region $i$ $(i=1,2)$
as $b_i(D)$, 
\be
b_1(D)=-\frac{D}{2}\coth DM\;,\qquad
b_2(D)=-\frac{D}{2}\cot DM\;. 
\ee 
Now using Eq. (\ref{power21}) we have an expression for the radiated power as
\be\label{fg}
P=\frac{(2E)^{3/2}}{3\pi}
\{\int^{\infty}_0dD\frac{(2a_1b_1''-a_1'b_1')^2}{4a_1^{7/2}}+
\int^{\pi/M}_0dD\frac{(2a_2b_2''-a_2'b_2')^2}{4a_2^{7/2}}\}\;,
\ee
again $a_1(D)$, $a_2(D)$ are the metric coefficients in each region, 
$E$ is the conserved energy, and $a_1'$ denotes the derivative of $a_1$ 
with respect to $D$ etc. The two integrals correspond respectively to the 
radiation produced from the two sections of the geodesic. Both integrals 
can be computed numerically, their sum has the approximate behavior
\be\label{slt}
P\approx\frac{E^{3/2}}{12\pi(2M)^{1/2}}
\{\frac{0.01}{(0.3+m/M)^{7/2}}+
4.3\sqrt{\frac{M}{m}}\}\;.
\ee
Holding $M$ fixed we clearly see that the second term above diverges 
as $m\rightarrow 0$. The divergence comes from Region 2,
where the $\betabf$ monopole separates from the donut configuration of two 
$\alphabf$ monopoles. The divergence arises as $D$ approaches $\pi/M$, 
i.e. as the $\betabf$ monopole approaches infinity. Although we have only 
calculated the radiation from a single geodesic we expect to encounter a 
similar behavior for any geodesic in which the $\betabf$ monopole is 
asymptotically well separated from the two $\alphabf$ monopoles.
As mentioned earlier we should really work in the overall center of mass 
frame. This has the effect of changing $m$ to the reduced mass, and 
multiplying the overall result by a function of $m/M$. We have omitted
this correction term for simplicity, as $m\rightarrow 0$ 
the correction becomes negligible.

To gain some insight as to the source of the divergence in Eq. (\ref{slt})
we examine more closely the monopole dynamics during the above scattering 
event. We restrict our attention to Region 2, where the divergence
arises. The metric in Region 2, Eq. (\ref{trigmetric}), has 
two terms, a $m$ dependent term and a term independent of $m$. 
The $m$ dependent term [the second line in  Eq. (\ref{trigmetric})]
can be written as $\frac{1}{2}\,m\,\dot{{\bf r}}_{\beta}^2$, with 
${\bf r}_{\beta}$ given by (\ref{rtrig}), it 
describes the kinetic energy of the $\betabf$ monopole  
positioned at ${\bf r}_{\beta}$. We call this the mass 
term, since it describes the extended particle-like behavior of a massive 
soliton. When $m$ is taken to zero the static energy density is not 
concentrated around the position of the $\betabf$ monopole, all that 
remains of the $\betabf$ monopole is a cloud surrounding the massive 
$\alphabf$ monopoles. The term independent of $m$ describes the cloud 
dynamics in the massless case, we denote this term as the cloud term.

Holding $M$ fixed, if $m>M$, the mass term is greater than the cloud 
term for all values of $D$ (or $|{\bf r}_{\beta}|$).
We interpret this as meaning that the configuration is composed of  
the donut $\alphabf$ monopole and a well defined $\betabf$ monopole 
(its energy density localized around its position). The 
cloud term in the Lagrangian describes the interaction of 
the $\betabf$ monopole with the donut $\alphabf$ monopole.
If $m<M$, the cloud term is greater than the mass term 
for $D$ less than some $m$ dependent value, $D_m$, which is determined from 
Eq. (\ref{trigmetric}). Since from Eq. (\ref{rtrig}),
$|{\bf r}_{\beta}|$ is an increasing function of $D$, 
then the cloud term is greater than the mass term 
for $|{\bf r}_{\beta}|$ less than some value $r_m$, roughly 
given as $r_m\approx 1/m$. This is so because for large 
$|{\bf r}_{\beta}|$,  Eq. ({\ref{trigmetric}) is approximately  given by
\be
L\approx\frac{1}{2}(m+\frac{1}{|{\bf r}_{\beta}|})\,\dot{{\bf r}}_{\beta}^2\;.
\ee
For $|{\bf r}_{\beta}|> r_m$ the mass term is greater than the cloud
term, as $|{\bf r}_{\beta}|\rightarrow\infty$ the mass term dominates.
For $|{\bf r}_{\beta}|\gg r_m$ the configuration should regain its 
interpretation as a distinct $\betabf$ and donut $\alphabf$ monopole. 
As $m\rightarrow 0$ with  $|{\bf r}_{\beta}|\ll r_m$ the cloud term 
dominates the kinetic energy of the configuration. Here, from analogy to 
the $m=0$ case, we expect the configuration to look like the donut 
$\alphabf$ monopole surrounded by some form of cloud configuration.
Previously, the term cloud has been used only in the massless limit, 
there it denotes the region in space inside which the monopole fields 
do not commute with all the generators of the non-Abelian unbroken gauge 
group. We will use the same term here as meaning the overall core size of the 
monopole. The cloud we discuss here becomes identical to the more familiar 
cloud in the massless limit. Inside the cloud the 
fields are non-Abelian and the overall core of the monopole is defined by the
cloud. We do not have explicit field information off the axis of symmetry so 
we cannot determine the nature of the cloud configuration and how it differs 
to that of $m=0$ case, \cite{I}. We expect the $\betabf$ monopole to appear 
as a distinct soliton only when $|{\bf r}_{\beta}|\gg r_m$ ($\approx1/m$). The 
value of $r_m$ increases to infinity as $m\rightarrow 0$. 

The apparent divergence in the radiation can be understood as follows. 
When the monopole cores overlap the monopoles lose their individuality. 
In Region 2, when the cores of the donut $\alphabf$ monopole and the 
$\betabf$ monopole overlap, the configuration resembles a cloud 
surrounding the donut $\alphabf$ monopole. The cloud radius is given by  
$|{\bf r}_{\beta}|$. Once ${\bf r}_{\beta}$ is far enough from the donut 
$\alphabf$ monopole, i.e. ${\bf r}_{\beta}\gg r_m$, the $\betabf$ 
monopole regains its individuality. This is as expected since the core size 
of a single $\betabf$ monopole is of order $1/m$, or $r_m$. 
By examining the radiation formula, Eq. (\ref{power21}), it can be seen that 
the radiation produced is significant for all $|{\bf r}_{\beta}|<r_m$. 
When $|{\bf r}_{\beta}|\gg r_m$ the monopoles are almost non-interacting 
(the metric is flat in terms of $|{\bf r}_{\beta}|$) and little radiation 
is produced. It is the cloud itself which is responsible for the diverging 
contribution to the radiation (the cloud is dynamical and moves with velocity
$\dot{\bf r}_{\beta}$). The donut $\alphabf$ monopole is almost static and is
not responsible for large amounts of radiation. As $m$ decreases towards 
zero, radiation is produced over a larger and larger time period and this is 
what causes the eventual divergence.  

With $M$ fixed and $m\ll M$ the above calculations imply that the geodesic 
approximation breaks down as the total radiation produced becomes of the 
same order as the kinetic energy, or $P\approx E\approx \frac{1}{4}M\,v^2$ 
where $v$ is the asymptotic relative velocity of the incoming $\alphabf$ and 
$\alphabf+\betabf$ monopoles. This occurs for incoming velocities $v$ of the 
order, $v^2\approx m/M$; the velocity of the outgoing $\betabf$ 
monopole, $\dot{{\bf r}}_{\beta}$, is of the order 1. In fact, holding the 
incoming velocity $v$ fixed and reducing $m$, it is easy to see that the 
outgoing velocity of the $\betabf$ monopole (considered as a function of $m$) 
increases without limit as $m\rightarrow 0$. So
even if the incoming monopole velocities are small, the velocity of
the outgoing $\betabf$ monopole becomes relativistic once its mass is small 
enough and the radiation produced correspondingly increases.

In the previous case of charge $(1,1)$ monopoles, as one of the 
masses is taken towards zero, the radiation produced remains finite. Both 
cases share the property that as one of the monopoles masses approaches 
zero the monopole core will increase to arbitrary large size. The difference 
lies in the fact that the monopole velocities remain small at all times 
for charge $(1,1)$ monopoles. For charge $(2,1)$ monopoles the velocity of the 
$(0,1)$ monopole becomes relativistic at small enough masses. This is what 
causes the large radiation above.    
 
The results above appear to indicate that the validity of the moduli space 
approximation will break down for charge $(2,1)$ monopoles as the mass of the 
charge $(0,1)$ monopole approaches zero. However we must first analyze more 
closely the assumptions made. The main assumption was that the 
monopole cores evolve according to the moduli space approximation. 
Using this, the dipole radiation of massless fields from the 
monopole core to spatial infinity was computed. 
Considering Region 2 of 
the above geodesic the monopole fields are non-Abelian inside the cloud 
(the overall core region). It is from the cloud to infinity that we 
calculated the radiation produced, in particular we determined the time 
dependent dipole moments outside the cloud where the fields are Abelian. 

We have also seen above that as the 
$\betabf$ monopole regains its individuality its asymptotic velocity 
increases without limit as a function of $m$ as $m\rightarrow 0$. 
The fields in the 
cloud region will also have very large time derivatives. Thus the moduli 
space approximation implies that the fields will have large 
time derivatives even if the incoming monopole velocities in Region 1 are 
small. Because the fields in the cloud region are highly relativistic it
is not justified to assume the the time dependent dipole moments outside 
the cloud are given by the geodesic approximation. To determine the time 
dependent dipole moments outside the cloud a proper consideration of the 
time dependent field equations is necessary. However this is a very difficult 
task. A proper inclusion of these effects will alter significantly the 
results found here for the dipole radiation produced in a scattering event. 
We know that the finite energy monopoles cannot actually radiate infinite 
energy, the question is whether or not their radiation is comparable to
their kinetic energy as given by the geodesic approximation. 
The true time dependent dipole 
moments will be very different to that predicted by the moduli space 
approximation, and using these correct dipole moments in Eq. (\ref{power})
may imply that the radiation produced is not too large. The only concrete
conclusion that we can draw is that the true time dependent monopole fields
in the cloud region will differ significantly from that predicted by the 
geodesic approximation.

We conclude by mentioning some possibilities for future work. 
It may be possible to make better progress by considering the SU(4) 
monopoles discussed in \cite{WY}, where explicit field information is known. 
As mentioned earlier the dipole moments change discontinuously when the 
unbroken symmetry group becomes non-Abelian. It would be useful to see how 
this occurs explicitly. It is likely that the divergence in the radiation 
found above is generic whenever there is a cloud configuration, again this 
can be tested by considering the monopoles charges in \cite{WY}.

It would also be helpful to repeat the calculation of Section 4 directly in 
the minimally broken theory. One must re-derive the radiation formula 
in terms of the multipole moments which are now non-Abelian.
This appears to be difficult because of the non-Abelian nature of the field 
equations. There are two massive monopoles with well defined heavy 
cores, parametrized by coordinates which appear in the moduli 
space. There are further moduli representing the cloud radius and its SU(2)
orientation. The same difficulty remains that for large cloud moduli 
there exist large regions in space where the fields are non-Abelian but 
the potential energy is small. The kinetic energy carried by the cloud 
parameters is however of the same order as that of the kinetic energy of the 
massive monopoles. This is what naively causes the divergence in the radiation
produced. Finally, it appears to us that a proper resolution of the 
questions raised here will require an analysis similar to that of
\cite{Stuart} for higher gauge groups.

\noindent{\bf Acknowledgments}
I thank Bernd Schroers for discussions which led to this work 
and Conor Houghton for helpful comments and a critical reading of 
the manuscript. I also thank FCAR of Qu\'{e}bec and NSERC of Canada for 
financial assistance.
\\[1cm]

\end{document}